# Platinum is a Photocatalyst: Large Visible-Light Quantum Efficiency Revealed

Olivier Henrotte[1,2*], Yin Chak Wong[3], Jordan Edwards[3], Simão Meneses João[3], Štěpán Kment[1,4], Emiliano Cortés[2], Johannes Lischner[3], Alberto Naldoni[5*]

1. Regional Centre of Advanced Technologies and Materials, Czech Advanced Technology and Research Institute, Palacky University, Olomouc, Slechtitelu 27, 77900 Olomouc, Czech Republic
2. Nanoinstitut München, Fakultät für Physik, Ludwig-Maximilians-Universität München, 80539 Munich, Germany
3. Department of Materials and Thomas Young Centre for Theory and Simulation of Materials, Imperial College London, SW7 2AZ London, United Kingdom.
4. Nanotechnology Centre, Centre for Energy and Environmental Technologies, VSB-Technical University of Ostrava, 17. listopadu 2172/15, 708 00, Ostrava, Poruba, Czech Republic
5. Department of Chemistry and NIS Centre, University of Turin, Turin 10125, Italy

e-mail: o.henrotte@lmu.de; alberto.naldoni@unito.it

ABSTRACT

Metal–semiconductor junctions in optoelectronic devices are commonly engineered to promote charge separation. In Pt/$TiO_2$ Schottky junctions, Pt is typically regarded as a catalytic electron sink rather than a visible-light-active component. Here, we demonstrate that Pt nanoislands on $TiO_2$ can generate photochemically active carriers under visible light excitation. Using quantitative scanning photoelectrochemical microscopy, we measure the wavelength-resolved external quantum efficiency (EQE) of Au and Pt nanoisland arrays on $TiO_2$, and correlate their reactivity with their morphology and extinction spectra. Discrete ~10 nm Pt nanoislands exhibit robust broadband visible light photoactivity—exceeding the photoactivity of similar-sized Au nanoislands under blue–green excitation—whereas Pt's photoactivity is strongly suppressed when the nanoislands are connected. Surprisingly, Pt exhibits an EQE-per-atom ~20 times higher than Au at 455 nm and ~2 times higher at 595 nm (at Au's optimum). We show an approximately wavelength-independent Pt internal quantum efficiency of ~1% across the visible spectral region. These findings reposition catalytic metals with strongly damped optical response in the visible as light-responsive components in metal-semiconductor hybrids, challenging the prevailing perception that they function solely as passive co-catalysts in photocatalytic systems.

## Introduction

Metal-semiconductor hybrid materials are central to photocatalysis and photoelectrochemistry because they combine strong optical absorption with efficient interfacial redox chemistry.[1,2] $TiO_2$ remains a benchmark semiconductor in this area owing to its stability and well-established surface science, yet its wide bandgap limits direct visible-light utilization. A common strategy is therefore to integrate metallic nanostructures that (i) harvest visible photons and generate highly energetic ("hot") carriers, or (ii) promote charge separation and catalytic turnover at the solid-liquid interface.[3] Gold (Au) is often used as the metallic light harvesting component: it supports pronounced localized surface plasmon resonances (LSPRs) in the visible range of the electromagnetic spectrum, enabling strong optical absorption and near-field concentration that can drive photoredox chemistry or hot-carrier injection into adjacent materials. In contrast, platinum (Pt) is commonly viewed as a co-catalyst and electron sink—an efficient site for catalytic reactions and a reservoir for photogenerated electrons originating from the semiconductor—rather than as an optical antenna or primary light absorber, such as Au.[4]

Following this classification, the community usually describes Au nanoparticles (NPs) on $TiO_2$ as exhibiting LSPR-enabled hot-carrier generation/near-field effects, with size-, geometry- and environment-dependent optical resonances producing characteristic wavelength-selective activity.[5] Pt-$TiO_2$ hybrids, by contrast, are often thought of as optically inefficient in the visible because Pt's resonances are strongly damped by substantial interband losses.[6,7] Moreover, Au and Pt exhibit a different hot carrier relaxation dynamics under visible-light excitation with charge carrier thermalization in Pt (described by the electron phonon-phonon coupling parameter,[8] $G_{ep}>10^{17}$ W m$^{-3}$ K$^{-1}$)[9] being one to two order of magnitudes faster than in Au ($G_{ep}=2.3\times10^{16}$ W m$^{-3}$ K$^{-1}$)[10], which means that electrons thermalize much faster in Pt than in Au.[11] This contrast is also reflected in their carrier transport lengths: the electronic mean free path in Au is comparatively long and can reach tens of nanometers (~40 nm),[12] whereas in Pt it is typically ~10 nm, consistent with the stronger electron scattering in Pt.[13] Pt's optical response in the visible, its strong electron-phonon coupling, and its short free carrier mean free path have led to the common expectation that its quantum efficiency of extracting photogenerated charge carriers under visible-light excitation is low.

However, this widely accepted classification is increasingly difficult to reconcile with a growing body of work showing that Pt nanostructures can exhibit appreciable visible-light extinction and participate in photoredox processes when coupled to high-index dielectric environments or engineered interfaces. A pivotal example is the demonstration that near-field dielectric scattering promotes the visible-light activity of Pt-hybrid nanostructures.[14–16] Additionally, a recent study demonstrated that Pt NPs are photoactivated by the strongly enhanced local electric field in the hot-spots of an Au NP supercrystal, with negligible influence from heat and charge transfer.[17] Moreover, quantitative measurements of absorption and scattering cross-sections of Pt and Au nanodisks show that Pt can be strongly absorption-dominated in the visible, with broad and heavily damped spectral features.[6,18] A recent kinetic study further indicate that Pt illumination can drive direct photochemical surface pathways.[19] Collectively, these studies suggest that the absence of a sharp LSPR peak does not mean that Pt cannot harvest visible light, but it indicates that Pt converts absorbed photons predominantly into electron-hole pairs.

The bottleneck towards establishing Pt as a photoactive material is that the field still lacks a quantitative, operando framework linking Pt's broadband visible absorption to chemically productive charge extraction at the solid-liquid interface. In other words, the presence of visible-light extinction does not guarantee photoactivity: electron-hole pairs must be separated and extracted at an interface before thermalization and recombination erase their chemical potential. This competition is especially severe for metals, where electronic relaxation is a fast process, necessitating interfacial extraction to occur on short time and length scales.[8] For Au-based hybrids, an intuitive structure-optics-reactivity relationship exists because LSPR tuning provides clear spectral markers and size trends. For Pt-based hybrids, the mechanistic narrative is fragmented: Pt is typically viewed as an electron sink, and studies emphasizing Pt's optical role often rely on specialized scatterer geometries or ensemble measurements that do not resolve where activity originates, how it depends on nanoscale morphology/connectivity, or why chemically productive charge extraction emerges in some Pt-$TiO_2$ systems but not others. Overall, this motivates a more careful investigation of hybrid materials under local operando conditions relevant to photoprocesses at the solid-liquid interface.

Here we address this gap by combining quantitative scanning photoelectrochemical microscopy (SPECM) with co-localized morphology and spectroscopy to directly compare Au and Pt nanoisland populations under identical working conditions on a $TiO_2$ thin film—namely Au-$TiO_2$ and Pt-$TiO_2$, respectively. Our approach decouples optical absorption from interfacial extraction by measuring the local external quantum efficiency (EQE) for a fast redox probe (ferrocene dimethanol; FcDM). We demonstrate that Pt nanoislands on $TiO_2$ can display strong visible-light-driven photoactivity if the metal is nanostructured into discrete, non-connected islands and supported by a $TiO_2$ film that enables efficient charge separation and extraction. By comparing Au and Pt nanoislands, we find that Pt can function as an effective visible-light absorber in the hybrid architecture, with efficiencies comparable to Au. The observed response is consistent with Pt's lossy optical behavior in the visible and with an interfacial energy level alignment that supports carrier separation at the liquid-metal-semiconductor junction. Collectively, these results identify Pt as an active participant in the photogeneration/extraction process in metal-$TiO_2$ hybrids, rather than a passive electron trap.

## Results and discussion

### SPECM platform and sample architecture

To enable a direct photoactivity comparison between Au and Pt, we performed SPECM measurements on co-fabricated Au and Pt nanoisland arrays prepared on the same $TiO_2$ support (Figure 1a). Au and Pt thin films were deposited as patterned micro-squares and subsequently transformed into nanoisland arrays by controlled thermal dewetting (see Section 1 in the Methods for fabrication and dewetting conditions). The methodology follows our previously reported approach, in which oxidized species generated under illumination are detected in solution by an ultramicroelectrode (UME; with an active radius $r_{tip}$ = 0.5 μm, and a ratio between the insulating part radius and active radius RG = 50).[20] Importantly, the substrate is electrically floating during all experiments; only the probe (UME) is biased for product detection. This configuration ensures that the recorded currents report on local interfacial photochemistry. The measured photoactivity ($\Delta I$) corresponds to the differential current (under illumination and in the dark).

Because the measured signal reflects the net outcome of photoexcitation—interfacial charge separation and carrier transfer to molecules in the electrolyte—we first define the working energetic framework for Au-$TiO_2$ and Pt-$TiO_2$ (Figure 1b). The expected behaviors of Au and Pt under visible light excitation can be explained by their electronic band structures and resulting optical loss channels near the Fermi level (Figure 1b-d). In Au, the filled 5d bands have an onset at ~2 eV below $E_F$, and Au's photoresponse depends on whether electron-hole pair excitations are predominantly intraband (i.e., involving only sp-band states) or interband (creating d-holes) in character. The onset of interband absorption in Au occurs near ~2 eV (620 nm), which can damp the plasmonic response and increase the fraction of absorbed energy that rapidly thermalizes electronically.[18,21] Because carrier extraction is an interfacial process, the Au NP size influences the measured photoactivity in multiple ways: via spectral tuning, but also because the non-equilibrium carrier generated within the metal needs to reach the interface before thermalization (typical non-equilibrium carrier diffusion lengths are on the order of tens of nanometers).[22,23]

In Pt, by contrast, the high density of d-states near $E_F$ produces strong damping and broadband absorption across the entire visible spectrum, leading to an optical response dominated by a broad set of interband transitions rather than a sharp, tunable LSPR. Measurements on nanodisks explicitly show that Pt is absorption-dominated with strongly damped plasmon-like collective oscillation compared to Au.[6,7] This suggests that visible-light illumination can generate electron-hole pairs in Pt efficiently. To distinguish the extracted carriers from the recombined ones, we consider effective electron Schottky barrier heights of $\Phi_B = 1.1$ eV for Au-$TiO_2$ and $\Phi_B = 1.3$ eV for Pt-$TiO_2$ (Figure 1b).[24] We emphasize that these values are representative parameters to support energy band-diagram construction and mechanistic discussion, as the effective Schottky barrier may vary at the nanoscale.[25] In light of these considerations, the central question is to what extent these excitations translate into chemically productive interfacial charge transfer, which is governed by the competition between charge carrier separation/extraction and thermalization/recombination.[8,9,11]

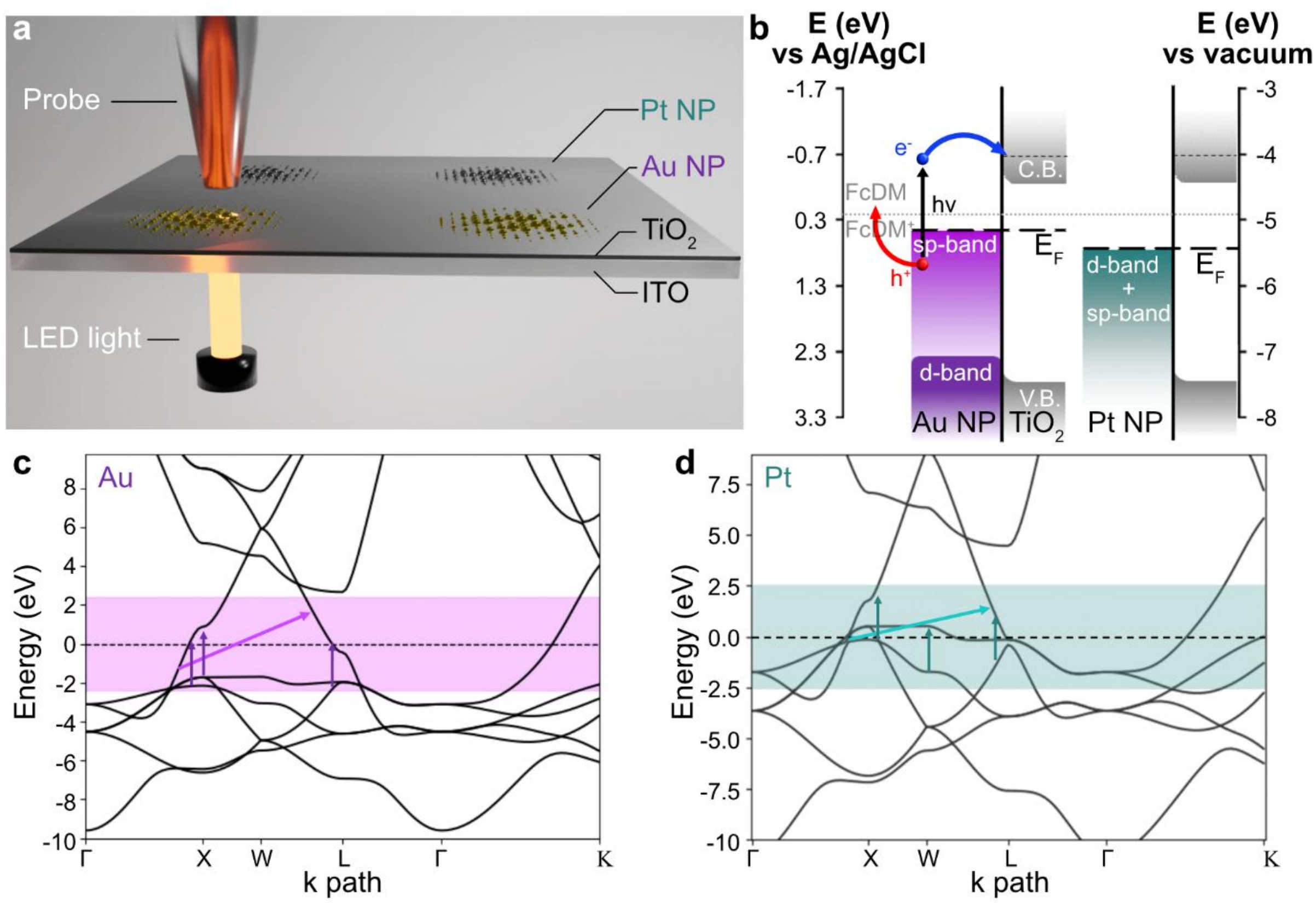

**Figure 1.** Scanning photoelectrochemical microscopy platform and energetic framework for Au-$TiO_2$ and Pt-$TiO_2$ photoactivity. (a) Schematic of the SPECM configuration. (b) Energy-level band showing the relevant electronic levels of Au and Pt (Fermi level $E_F$; schematic sp- and d-bands), the $TiO_2$ band edges (conduction band, C.B.; valence band, V.B.), and the FcDM/$FcDM^+$ redox couple. (c,d) Electronic band structure of Au (c), and Pt (d) from tight-binding highlighting representative optical excitation pathways under visible illumination.

**Wavelength-dependent photoactivity and EQE**

To address this question experimentally, we combined position-resolved SPECM with co-localized structural and optical characterization on the same nanoisland arrays. We first acquired a positioning map to identify the Au and Pt arrays and to define a consistent scan trajectory across the patterned regions (Figure 2a). We then recorded wavelength-resolved line scans across each array and converted the collected probe currents into an external quantum efficiency (EQE) using a diffusion model (Figures 2b & S1; see Sections 2 and 3 in the Methods).[26] Because the dewetting process generates systematic gradients in nanoisland size and connectivity across each array, the quantitative interpretation of Figure 2b requires assigning to each spatial position a nanoisland population and its optical response (Figure 2c-e).

To rationalize the wavelength- and position-dependent EQE trends, we quantified the nanoisland morphology as a function of lateral position within each array (Figure 2c,d; Figures S2-4). This analysis reveals distinct size regimes across the scan line and enables a controlled comparison between Au and Pt populations with similar characteristic dimensions. In the following, we focus on two representative populations corresponding to approximately 10 nm and 30 nm nanoisland arrays (Figures 2d & S4; Table S1). The diameters of the chosen population were (i) 34±11 nm and 11±7 nm with a surface coverage of 26.3% and 8.5%, respectively, for Au (Figure S2a,d); and (ii) 33±23 nm and 9±9 nm with a surface coverage of 26.5% and 9.8%, respectively, for Pt (Figure S3c,d).

We next obtained local extinction spectra at the positions corresponding to these populations using the SPECM platform coupled to spatially resolved optical characterization (Figure 2e; see Section 4 in the Methods).[26] Au exhibits a distinct LSPR feature that shifts to longer wavelength with increasing island size: ~540 nm for ~11 nm islands and ~600 nm for ~34 nm islands. In contrast, Pt nanoislands display broadband extinction across the visible, with a wide maximum spanning approximately 400–700 nm and modest size-dependent shifts of the maximum position: ~450 nm for ~9 nm islands and ~480 nm for ~33 nm islands. This spectral behavior is consistent with the strongly damped optical response of Pt, where extinction is predominantly absorptive and possible resonances are completely damped by optical losses.[6]

With the position-resolved nanoisland populations established, we now discuss the wavelength-dependent EQE line scans (Figure 2b). For the Au array, the EQE increases from 455 nm toward the green–orange region peaking at 595 nm (~0.06%) and then decreases toward the near-infrared, approaching negligible activity at 850 nm. Consistent with the size-dependent plasmonic response, the array center—dominated by larger Au nanoislands—exhibits a higher EQE at longer wavelengths (660 nm and higher) than the array edges, which contain smaller islands.[26] This trend aligns with the expected red-shift of the Au LSPR and the associated redistribution of optical absorption and near-field enhancement with increasing particle size (Figure 2d).[21]

Strikingly, the Pt array displays a qualitatively different spatial dependence. While the center of the Pt array shows almost no activity, the array edges exhibit a pronounced EQE that, under blue–green excitation (<550 nm), can match or exceed the response measured on Au under identical conditions with an EQE reaching ~0.09% at 455 nm. The suppression of activity at the Pt array center correlates with the onset of nanoisland connectivity (Figure S3), which likely introduces efficient recombination pathways. Importantly, additional Pt arrays with reduced connectivity exhibit substantial activity within the central region (Figures S5 and S6), reinforcing that Pt photoactivity is maximized for discrete islands rather than continuous metallic networks. Finally, Pt retains detectable activity even at 850 nm, consistent with its broadband extinction extending into the red/near-IR.[15]

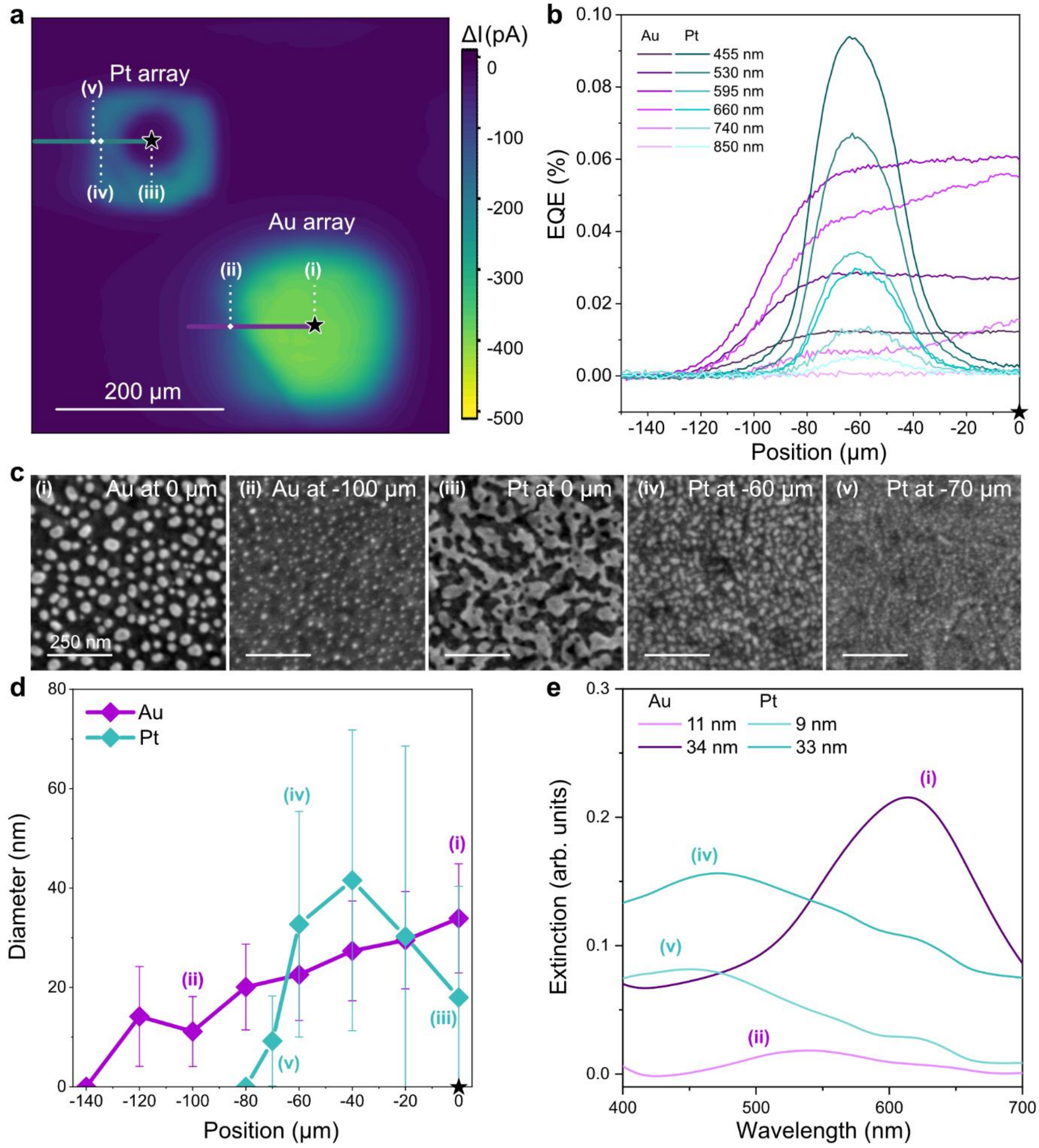

**Figure 2.** Position-resolved external quantum efficiency (EQE), morphology, and optical properties of Au and Pt nanoislands. (a) SPECM photoactivity map acquired over the Pt nanoisland array and the Au nanoisland array under visible illumination, used to identify the scan trajectories and regions of interest; the stars indicate the centre of the arrays. The markers (i) to (v) correspond to positions of interest. (b) EQE as a function of position along the scan line indicated in (a); 0 μm corresponds to the stars in (a). Au data are shown from purple to pink and Pt data from teal to cyan, with each trace corresponding to a different excitation wavelength. (c) SEM images of the Au array after dewetting at the line scan position of 0 μm (i) and -100 μm (ii), and of the Pt array after

dewetting at the line scan position of 0 μm (iii), -60 μm (iv), and -70 μm (v). (d) Position-resolved nanoisland diameters (filled diamonds) for Au (purple) and Pt (teal) extracted from SEM images for the same trajectories than (b). (e) Extinction spectra measured at positions corresponding to the nanoisland diameters identified in (d). SPECM measurements' conditions were $r_T$=5 μm, RG=5 and $v_{scan}$ = 50 μm/s for (a) and $r_T$=0.5 μm, RG=50 and $v_{scan}$ = 20 μm/s for (b). The measurements were performed in an aerated solution of 1 mM $Fc(MeOH)_2$ and 0.1 M KCl at Z=10 μm.

**Hot carrier generation and extraction at the interface**

Motivated by the unexpectedly high EQE observed for Pt NPs, we analyze our experimental findings by theoretically investigating environment-dependent hot-carrier generation in Au and Pt NPs. Specifically, we perform atomistic calculations of hot-carrier generation rates in Pt and Au nanoparticles using the approach developed in Ref. [27] (see Section 5 in the Methods). This is essential because Pt is still commonly discussed as an electron sink that improves charge separation in $TiO_2$, rather than as a visible-light absorber on its own—even in recent works.[4,28,29]

The dielectric environment affects hot-carrier processes primarily by (i) modifying the optical mode energy and field distribution (and thus the balance between intraband and interband excitation) and (ii) regulating the carrier yield through environmental screening of the electron-photon interaction. Recent studies show that, in plasmonic nanostructures, increasing the dielectric constant of the surrounding medium can suppress hot carriers generated via interband transitions while enhancing intraband-derived hot-carrier generation, highlighting the dielectric environment as a design parameter on the same footing as material choice and particle size.[27,30] Consistent with this framework, we calculated the hot-carrier generation rate under specific wavelength excitation for Au and Pt nanoislands, explicitly accounting for nanoparticle size and the optical dielectric environment imposed by $TiO_2$ (Figure 3, see Section 5 in the Methods).

To connect the model to our experimental results, we define the generation rate of potentially extractable holes. This rate is assumed equal to the generation rate of hot electrons with energies exceeding $\Phi_B$ at the metal-$TiO_2$ junction, as electrons injected into the $TiO_2$ cannot recombine with holes in the metal NP resulting in long-lived holes that can be extracted at the NP-electrolyte interface. We note that our hot carrier model does not explicitly include interfacial electronic states formed at the metal-$TiO_2$ junction.[20,31–33] Those interfacial states can introduce nanoscale variations in charge-extraction probability and effective Schottky barrier energy. Accordingly, this model does not take into account thermalization losses, nanoscale barrier inhomogeneity, and defect-mediated recombination but it provides a physically transparent metric and an upper limit for comparing how composition and dielectric environment affect photoactivity.

For Au, the calculated hot-carrier distribution is strongly dependent on the LSPR position. Under higher-energy excitation (shorter wavelengths), Au's absorption is dominated by interband transitions, which generate energetic holes but do not yield a large population of electrons with energies exceeding $\Phi_B$. This manifests as a comparatively smaller above-$\Phi_B$ hot electron population in the blue-green regime (Figure 3a), consistent with the reduced EQE observed experimentally (Figure 2b).

By contrast, the calculated population of electrons above-$\Phi_B$ for Pt increases monotonically with photon energy across the studied range (Figure 3b). This trend is consistent with Pt's broadband, strongly damped optical response in the visible (Figure 2e), and qualitatively mirrors the wavelength dependence of the experimentally collected EQE (Figure 2b). This correspondence supports the hypothesis that for Pt nanoislands the spectral dependence of carrier extraction is more directly linked to broadband optical absorption, mean free path, and barrier-filtered carrier generation than to a resonant plasmonic mode.

To further isolate the role of the environment, we compared the calculated hot-carrier generation rates for identical nanoislands embedded in two dielectric media: a high-permittivity environment representative of $TiO_2$ ($\varepsilon_m$=4; Figure 3) and a lower-permittivity one representative of a colloidal system (in water) or on a glass-like surrounding ($\varepsilon_m$=2; Figure S7). This controlled comparison allows us to evaluate how dielectric screening alters the predicted above-$\Phi_B$ carrier population and thus provides a mechanistic bridge between the simulated generation trends and the experimentally observed wavelength-dependent EQE (Figure S8, the choice of $\varepsilon_m$ is explained in Figure S8 caption).

For Au, increasing $\varepsilon_m$ produces the expected perturbation of the plasmonic response, consistent with the literature.[27,30] At shorter wavelengths (<500 nm), the predicted change in above-$\Phi_B$ electron population is comparatively small. In comparison, Pt shows a monotonic enhancement in above-$\Phi_B$ carrier generation as $\varepsilon_m$ increases, particularly in the blue-green region. For example, at 450 nm, the model yields a ~4-fold increase when $\varepsilon_m$ is raised from 2 to 4 (Figure S8). This trend supports an important message: Pt's photoresponse is more strongly modulated by the dielectric environment than Au's in the blue-green spectral region.

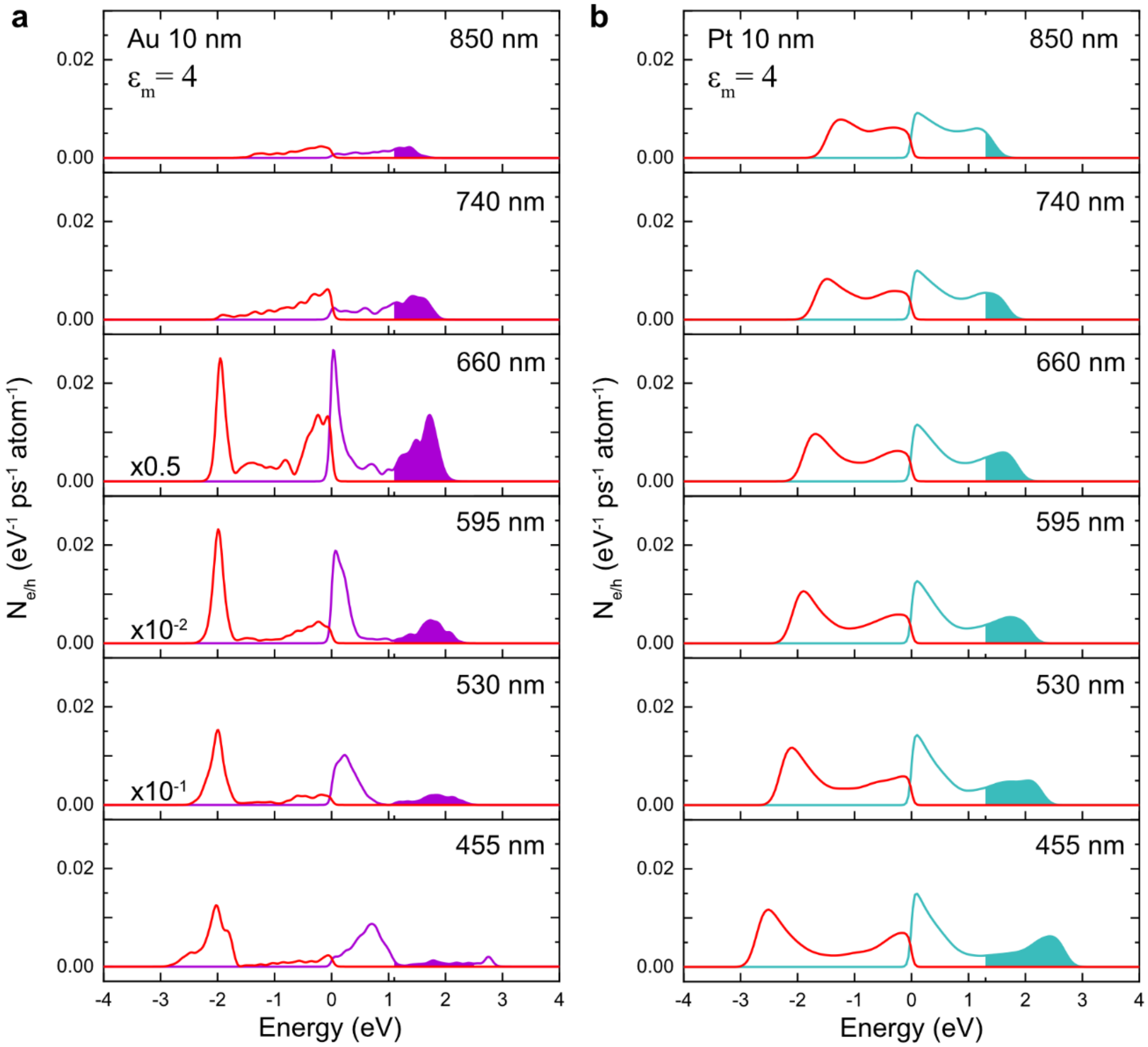

**Figure 3.** Theoretical simulation of hot carrier generation in Au and Pt nanoislands on a $TiO_2$ thin film. (a) Energetic distribution of hot electrons (purple line) and hot holes (red line) in a spherical Au nanoparticle of 10 nm diameter for different excitation wavelengths in a high-permittivity environment ($\varepsilon_m$=4). (b) Energetic distribution of hot electrons (cyan line) and hot holes (red line) in a spherical Pt nanoparticle of 10 nm diameter for different excitation wavelengths in a high-permittivity environment ($\varepsilon_m$=4). The number of above-Schottky barrier electrons (corresponding to the number of extractable holes) are represented by the filled region in purple for Au and cyan for Pt. The Fermi energy is set to 0.

Keeping in mind the hot-carrier generation calculations, we examine how the measured EQE depends on photon energy for the Au and Pt nanoisland populations (Figure S9), and how this dependence evolves with nanoisland size and surface coverage.

For Au nanoislands, the EQE exhibits a pronounced maximum in the orange-red spectral region (595 nm and 660 nm), i.e., near the LSPR of the larger nanoislands, and falls off rapidly towards higher and lower photon energies (blue-green and near-IR regions). In our data, the peak EQE for the 34 nm Au population reaches ~0.06%, about 4-times higher than the 11 nm population (~0.015%) under identical conditions (Figure S9a). An intermediate Au population (~20 nm) follows the same size-dependent trend (Figure S9a). This behavior is qualitatively consistent with the well-established optical response of Au nanostructures, where (i) the spectral position of the LSPR red-shifts with increasing size (Figure 2e), and (ii) the growing contribution of interband transitions at photon energies above the interband onset (~2 eV) increases the quantity of carriers that recombine before reacting.[34,35]

Pt nanoislands display a markedly different spectral response. For both 9 nm and 33 nm Pt populations, the EQE is highest at the shortest wavelength measured (455 nm, with EQE up to ~0.09%) and decreases monotonically toward the near-IR, while remaining measurable even at 850 nm (Figure S9b). A larger Pt population (42 nm) exhibited 3-times lower EQE at 455 nm compared with the smaller Pt islands (Figure S9b). This broadband behavior is consistent with the heavily damped optical response of Pt in the visible (Figure 2e), where absorption via electron-hole pair excitation dominates and resonances are broadened by strong electronic losses.[7]

Notably, the smallest nanoislands cover substantially less area than larger nanoislands (around 3 times less, see Table S1). Therefore, the obtained EQE combines intrinsic efficiency with how much metal is actually present and accessible. To separate these effects, we normalized the response using different EQE descriptors: (i) per nanoisland—using the island density from SEM images (Figure S10), (ii) per atom (considering the islands as perfect spheres) (Figure S11), (iii) per surface area (considering the surface of the nanoislands from perfect spheres) (Figure S12), and the outer-shell atom (considering 2 nm of the outer volume from perfect spheres)[36] (Figure S13). A rationale for the choice of normalization is provided in Supplementary Section 1. We use the per-atom metric as the main descriptor because it best reflects metal-utilization efficiency in a photoinduced charge-extraction process, where the nanoisland volume contributes to light absorption and carrier generation before extraction occurs at the interface. This normalization does not imply that all atoms are catalytically equivalent or equally accessible to the electrolyte; instead, it provides a consistent scaling parameter that accounts for differences in metal loading, surface coverage, and nanoisland size.

Using this normalization, the smallest Pt nanoislands remain strikingly efficient: at 455 nm, the Pt (9 nm) population exhibits an EQE-per-atom ~15-folds higher than Au (11 nm), while still ~1.2-times higher at 595 nm, which correspond to Au's optimum (Figure 4a). This demonstrates that, under the present conditions, small Pt nanoislands transduce absorbed photons into reacting charges with unexpectedly high effectiveness. Interestingly, the large Pt nanoislands (~33 nm) showed a 1.5-fold higher EQE than the small Au nanoislands (~11 nm) at 455 nm even after normalization (Figure 4a), highlighting the excellent photocatalytic performance of Pt under blue light.

When we compare the EQE with (i) the measured extinction (Figure 2e) and (ii) the simulated rate of above-barrier hot-carrier generation (Figures 4b & S8), Pt shows a close spectral correspondence across the visible range, whereas Au displays a modest mismatch around its LSPR maximum (Figures 4 & 2e). The Au deviation is consistent with the well-known sensitivity of Au plasmonic response to interband damping.[37] For Pt, the closer correspondence supports that optical absorption is a reasonable predictor of extractable carriers—but only when the interface and environment enable efficient separation, keeping in mind Pt electron mean free path of ~10 nm.

To further compare Au and Pt, we estimate an internal quantum efficiency (IQE) as EQE/A (λ), where A is absorption. Because absorption is difficult to measure directly for our smallest nanoislands, we approximate it from the extinction, while considering existing literature for similar systems. Pt nanodisks in the regime of tens of nanometers are reported to be strongly absorption-dominated, with negligible scattering in the visible. For Au, the scattering becomes increasingly important with size, but for sizes comparable to 30 nm in diameter, the scattering fraction at the LSPR remain on the order of ~10%.[6,7,37] Accordingly, we treat Pt as non-scattering and Au as low-scattering (~10%). We provided in the Supplementary Information an example of our data set for different scattering considerations (Pt for 0 and 10%, and Au for 0, 10 and 20%), which highlights the negligible influence of the scattering on our findings (Figure S14). As depicted in Figure S14, the ~30 nm nanoislands yield higher IQE for Pt than Au at 455 nm (0.58% vs 0.18%), while at 595 nm Pt and Au become comparable (~0.3%), reflecting opposite spectral trends: Pt IQE decreases toward longer wavelengths, whereas Au IQE increases moving away from the interband-damped region toward the more purely intraband regime.

For the ~10 nm nanoislands, Pt shows an approximately wavelength-independent IQE near ~1% across the visible; consistent with the idea that Pt's broadband absorption can translate into a relatively uniform absorption-to-carrier separation efficiency when the interfacial energy level alignment is favorable. Au, by contrast, reaches higher IQE (~2.4%) at the blue edge in our estimate, but shows a pronounced dip near the LSPR maximum (~0.6%)—consistent with the general expectation that the LSPR maximum does not necessarily maximize extractable carriers because ultrafast thermalization and competing decay pathways can increase at conditions where optical losses are strongest.[26,34,35,37]

Our findings confirm that Au remains an excellent resonant photocatalyst whose apparent advantage depends strongly on size-tuned spectral overlap and extraction/recombination competition, whereas Pt—despite being non-plasmonic—can act as a highly effective broadband light-to-carrier extraction material when structured at dimensions comparable to the carrier mean free path.

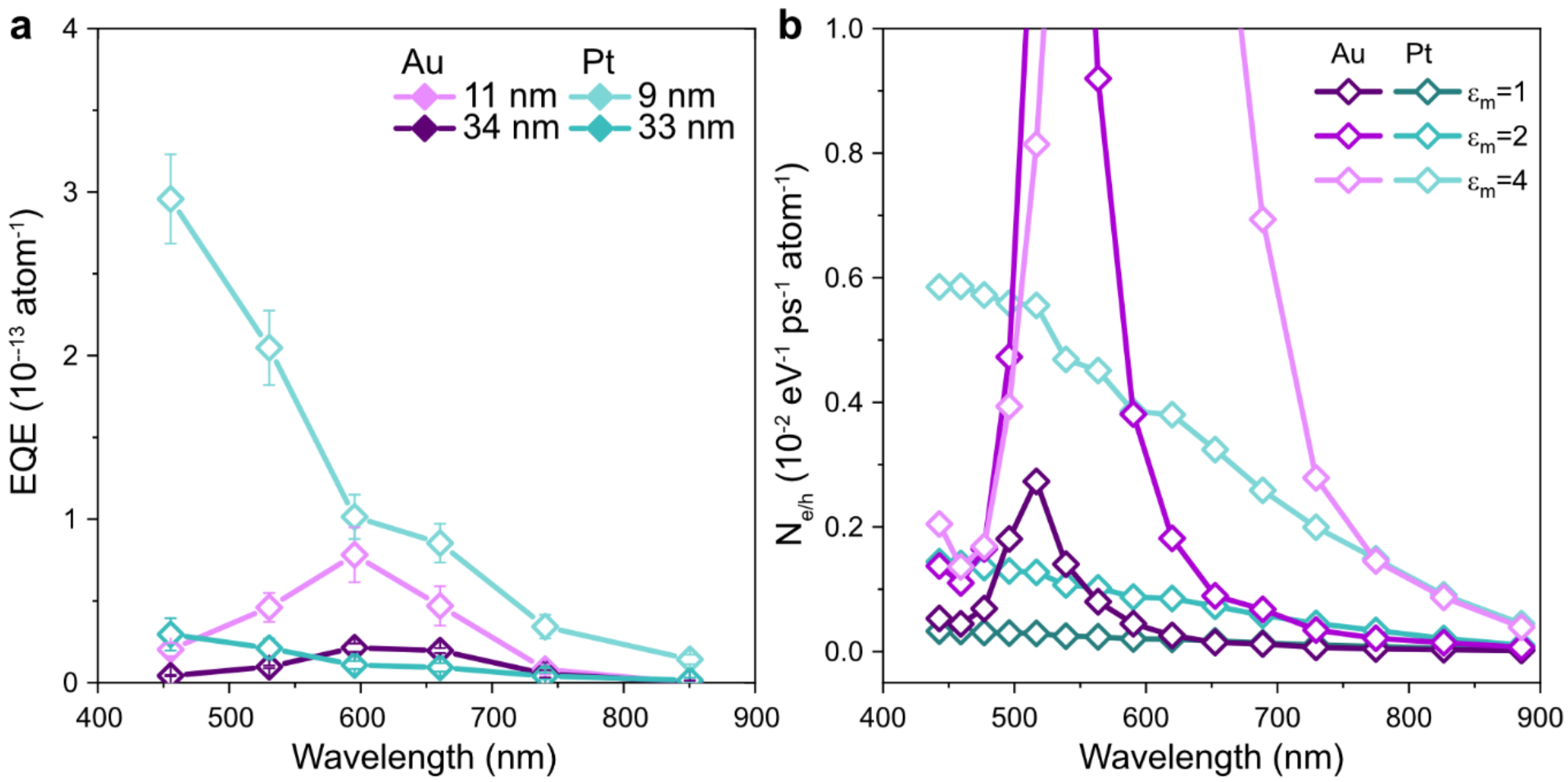

**Figure 4.** Comparison of external quantum efficiency and theoretical hot carrier generation with above Schottky barrier energy of Au and Pt nanoislands on $TiO_2$ thin film. (a) External quantum efficiency (EQE) per atom as a function of excitation wavelength for different populations of Au (11 nm, pink; 34 nm, purple) and Pt (9 nm, cyan; 33 nm, teal) nanoislands. (b) Calculated number of generated hot electrons above Schottky barrier per atom for a 10 nm nanoparticle of Au (purple to pink) and Pt (teal to cyan) for different $\varepsilon_m$.

## Conclusions

This work demonstrates via operando SPECM and theoretical investigations that charge carriers generated in plasmonic and catalytic metals on $TiO_2$ thin films under visible-light excitation can react with molecules in the surrounding media with similar efficiencies. We revealed a significant photoactivity for Pt at a distinct regime: discrete ~10 nm Pt nanoislands (corresponding to the electron mean free path in Pt) exhibit robust broadband visible-light activity—approaching or exceeding size-matched Au under blue-green excitation (EQE-per-atom 20-folds higher). Furthermore, Pt photoactivity is strongly suppressed when islands become connected, identifying connectivity as a critical parameter controlling chemical productivity. In contrast, Au shows the expected size- and wavelength-selective behavior associated with resonant plasmonic response. Additionally, the IQE is approximately wavelength-independent for Pt across the visible (~1%), consistent with broadband absorption translating into a relatively uniform absorption-to-extractable-carrier efficiency.

Overall, these findings address a key gap in metal-semiconductor photocatalysis: visible extinction in Pt is not intrinsically photochemically silent, but its conversion into interfacial redox activity depends sensitively on nanoscale morphology and environment conditions. Beyond repositioning Pt as an active participant in light-to-chemical energy conversion, our work suggests that transition metals, commonly adopted for catalysis, exhibiting strongly damped optical response in the visible are actively participating in photochemical processes. Controlling the local environment and interfacial junction at metal nanostructures is essential for maximizing photocatalytic performance. In this context, nanoscale operando tools, such as SPECM, offer a powerful strategy to provide the mechanistic insight needed to rationally design more efficient solar-to-chemical energy conversion architectures.

## Methods

### Section 1 – Fabrication and morphology of Au and Pt nanoislands on $TiO_2$.

**Fabrication**. The Au and Pt nanoisland arrays on $TiO_2$ thin films were prepared on 25x25 mm² ITO coated glass (Ossila, UK). The ITO coated glass substrates were cleaned by ultrasonication in acetone, ethanol then deionized water for 5 minutes each. A $TiO_2$ layer of 10 nm was deposited by plasma-assisted atomic layer deposition (ALD, Ultratech/CambridgeNanoTech Fiji 200) using tetrakis(dimethylamido)titanium as precursor (99% purity, Strem, USA) at 250 °C. An ALD cycle consisted of 0.1 s of precursor exposure, 5 s purge, 20 s $O_2$ plasma exposure (300 W), and 5 s purge of reaction products. The so-obtained $TiO_2$ thin layers were annealed at 450°C (2°C/min) under air for 2 h to insure $TiO_2$ anatase phase formation. A 10 nm thick Pt film was sputtered through a patterned mask (Micronlaser Technology, USA) divided into 16x16 squares (around 100x100 µm² spaced by 400 µm). The samples were then annealed at 550°C (2°C/min) under air for 2 h forming the Pt nanoisland arrays. Subsequently, the Au nanoisland arrays were formed upon annealing in air at 400°C (2°C/min) for 1 h of a 10 nm thick Au film sputtered through a similar mask on another region of the sample. This ensured well-defined and mono-elemental arrays of nanoisland gradients for both Pt and Au.

**Morphological characterization.** A field emission scanning electron microscope (FE-SEM, Hitachi SU 6600, Japan) was used to investigate the nanomaterials morphology. The Pt and Au size distribution was obtained by ImageJ applying a filter depending of the observed contrast. The calculated diameter were retrieved from the area of the nanoislands measured by ImageJ by approximating all nanoislands shape to a circumference.

### Section 2 – Scanning photoelectrochemical microscopy (SPECM).

SPECM experiments were conducted using an ELP 3 SPECM-FL (HEKA Elektronik GmbH, Lambrecht, Germany), a PG 618 USB bipotentiostat (HEKA Elektronik GmbH, Lambrecht, Germany) and fiber-coupled LED with different wavelengths: 455 ± 7, 530 ± 15, 595 ± 40, 660 ± 9, 740 ± 11 and 850 ± 15 nm (Thorlabs, USA). The SPECM employed an optical path below the sample to focus the LED light beam (~ 5 µm radius) with a 40x objective lens (LUCPLFLN 40x, NA = 0.6, Olympus, Japan) to locally irradiate the sample and drive the photochemical reactions, following a previously established methodology.[26] The experiments were performed in aerated solution of 0.1 M KCl and 1 mM $Fc(MeOH)_2$ (ferrocene dimethanol, $FeC_{12}H_{14}O_2$). Chemicals were bought from Sigma Aldrich with the highest purity and used without further purification. Milli-Q water was used for the preparation of all aqueous solutions. SPECM consists of a four-electrode setup with two working electrodes (WE): the substrate (WE1, not connected in this study) and the ultramicroelectrode (WE2), a Ag/AgCl wire used as reference electrode and a Pt wire as counter electrode. The ultramicroelectrode (from HEKA Elektronik GmbH, Lambrecht, Germany) used in this work has: $r_T$ = 5 µm and RG = 5 for the photoactivity position map (Figure 2a); and $r_T$ = 0.5 µm and RG = 50 for the EQE measurements (Figures 2b and 4a,b, and Figures S1, S5, S9, S10, S11, and S13). The probe moves in the vicinity (Z=10 µm) of the studied sample to detect the local species concentration in solution. The light beam and the probe were aligned before measurements. The samples were scanned on the XY stage while the probe was placed at a constant height (Z) using the Z stage. The redox mediator is oxidized from the extracted holes as follow:

$$Fc(MeOH)_2 + h^+ \rightarrow Fc^+(MeOH)_2 \qquad (1)$$

This locally modifies the concentration of both species (reduced and oxidized) in the vicinity of the beam position. The probe detects this reaction by reducing back (substrate generation – tip collection, SG-TC mode) the species produced by the investigated surface:

$$Fc^+(MeOH)_2 + e^- \rightarrow Fc(MeOH)_2 \qquad (2)$$

The probe is biased at a potential ($E_T$) corresponding to the diffusion limited regime (-0.1 V vs Ag/AgCl). Thus, the probe current ($I_T$) corresponds to a steady-state regime and is given by equation 3 in solution ($I_{T,\infty}$) and equation 4 near the substrate ($I_{T,ins}$) due to the hindering effect on the diffusion of both the substrate and the probe (corresponding to negative feedback):[38]

$$I_{T,\infty} = 4nFCDr_T \qquad (3)$$
$$I_{T,ins} = 4nFCDr_T\beta(RG)Ni_T(L,RG) \qquad (4)$$

where n is the number of electrons involved; F, the faraday constant (96485.33 C/mol); C, the concentration in species; D, the diffusion coefficient of the species ($7.8 \times 10^{-6}$ cm²/s); $r_T$, the probe active part radius; RG, the ratio between the probe insulating part radius and $r_T$; $\beta(RG)$, the RG impact on the diffusion to the probe active area and $Ni_T(L)$, the hindering effect of the near surface on the probe current with $L$ corresponding to the ratio between the probe-substrate distance ($Z$) and $r_T$.

The photochemical map acquisition was done with an XY scan method consisting of 1001 acquisitions in X and 51 lines in Y. The probe was scanning at determined scan speed ($v_{scan}$). The Z axis position was constant through the scan and was determined by approach curves nearby the zone of interest thanks to the negative feedback observed due to the $TiO_2$ insulating properties.

**Section 3 – Diffusion model.**

The diffusion model was simulated with COMSOL Multiphysics using the "Transport of Diluted Species" physics in a 2D asymmetric mode considers the substrate as an electrochemical object corresponding to the light beam size.[26] This simulation model gives information such as the ratio between the probe current and the substrate current (called collection efficiency, CE, equation 5), the probe geometry influence and the impact of the probe-substrate distance on the CE. The diffusion profiles were already provided in previous works.[23,26] CE refers to the ability of the probe to collect species produced by the substrate and is defined as the probe current divided by the substrate current:

$$CE(\%) = \frac{I_T}{I_{Sub}} \times 100 \qquad (5)$$

To avoid consequential influence from the probe on the measurements, every measurement were done with Z ≥ 10 µm. From the obtained CE, $I_{Sub}$ was calculated thanks to the measured ΔI. The external quantum efficiency (EQE) was calculated from the following equation:

$$EQE\ (\%) = \frac{I_{Sub}/e}{P/h\nu} \times 100 \qquad (6)$$

where e is the charge of one electron, P is the light power and hν is the photon energy.

**Section 4 – Scanning spectrophotometer microscopy (SSM).**

The optical properties were measured with a spectrophotometer coupled with the scanning photoelectrochemical microscope position system to locally investigate the substrate (HEKA Gmbh, Lambrecht, Germany). SSM consists of a white light focused through an objective (UPLFLN 60x, NA = 0.9, Olympus, Japan) below the sample and aligned with the spectrophotometer thanks to a glass capillary (100 µm radius) connected to an optical fiber (100 µm radius, NA = 0.22, Thorlabs, USA) above it to collect the transmitted light. The illumination beam radius was ~ 5 µm and the lamp used was a Lambda LS Xenon Arc bulb (wavelength from 330 to 670 nm, Sutter instruments, USA). The measured linescans were generated thanks to the localized transmission spectra at different XY positions of the sample (every 10 µm). From those spectra, the values from the wavelength of interest were extracted and combined together depending of their XY position to form the measured linescans. Those values correspond to $-Log\ (T)$, where $T = \frac{I}{I_0}$, with $I$ and $I_0$ corresponding to the transmitted light through the measured sample and the bare $TiO_2$/ITO substrate, respectively. Due to the white lamp limitation in wavelength and with lower light intensities at shorter and longer wavelengths, we were able to investigate wavelengths from 400 to 670 nm, with higher intensity (signal-to-noise ratio) between 475 nm and 600 nm.

**Section 5 - Hot-carrier modelling.**

To study hot-carrier properties of Pt and Au nanoparticles, we evaluate Fermi's golden rule using the MaxTB code.[27,39] We first construct atomistic models for spherical Pt and Au nanoparticles. For the Au nanoparticles, the electronic Hamiltonian is parametrized using the tight-binding model from Ref. [40] and for the Pt nanoparticles, the tight-binding model from Ref. [41] is used. The perturbing potential caused by the illumination is evaluated using the quasi-static approximation with experimental dielectric functions from Ref. [42].

**Data Availability Statement**

The data that support the findings of this study are available from the corresponding author upon reasonable request.

**Supporting Information**

The Supporting Information is available:

The Supporting Information consists of additional SPECM measurements, SEM images, and theoretical simulation of charge carrier generations.

**Author contributions**

O.H. conceived and designed the project. A.N. coordinated, supervised and financed the project. O.H. performed the samples preparation, SPECM measurements, and analyzed the data. Y.C.W., J.E., S.M.J. and J.L. performed the hot carrier generation theory and DFT calculations. O.H. prepared the figures and wrote the first draft. All authors discussed and edited the manuscript.

**Acknowledgements**

The authors acknowledge the support of the Czech Science Foundation (GACR) through the award n. 22-26416S. The authors thank Dr. Luca Mascaretti for the $TiO_2$ atomic layer deposition (ALD) and Jiří Hošek for the SEM measurements. ALD experiments were performed at CEITEC Nano supported by the CzechNanoLab Research Infrastructure (ID 90251) funded by MEYS CR. A.N. acknowledges the support from the Project CH4.0 under the MIUR program "Dipartimenti di Eccellenza 2023-2027" (CUP: D13C2200352001). O.H. acknowledges financial support from the European Union's Horizon Europe research and innovation programme under the Marie Skłodowska-Curie grant agreement No 101205585. We acknowledge funding and support from the Deutsche Forschungsgemeinschaft (DFG, German Research Foundation) under Germany´s Excellence Strategy – EXC 2089/2–390776260, the Bavarian program Solar Technologies Go Hybrid (SolTech) and the Center for NanoScience (CeNS). J.L. acknowledges funding from the EPSRC programme grant EP/W017075/1. Š.K. acknowledges support from the ERDF/ESF project TECHSCALE (No. CZ.02.01.01/00/22_008/0004587). This work was also supported by the European Union under the REFRESH – Research Excellence for Region Sustainability and High-tech Industries project (No. CZ.10.03.01/00/22_003/0000048) through the Operational Programme Just Transition.

**Competing interests**

The authors declare no competing interests.